\documentclass[twocolumn,showpacs,preprintnumbers,amsmath,amssymb]{revtex4}

\usepackage{graphicx}
\usepackage{dcolumn}
\usepackage{bm}

\begin{document}


\title{Exactly-solvable models of proton and neutron interacting bosons}

\author{S. Lerma H.$^{1}$, B. Errea$^{1}$, J. Dukelsky$^{1}$, S. Pittel$^{1,2}$ and P. Van Isacker$^{3}$}

\address{$^{1}$ Instituto de Estructura de la Materia, CSIC, Serrano
123, 28006 Madrid, Spain \\
$^{2}$ Bartol Research Institute, University of Delaware,
Newark, Delaware 19716, USA \\
$^{3}$ Grand Acc\'{e}l\'{e}rateur National d'Ions Lourds, BP
55027, F-14076 Caen Cedex 5, France}

\date{\today}

\begin{abstract}
We describe a class of exactly-solvable models of interacting
bosons based on the algebra SO(3,2). Each copy of the algebra
represents a system of neutron and proton bosons in a given
bosonic level interacting via a pairing interaction. The model
that includes $s$ and $d$ bosons is a specific realization of the
IBM2, restricted to the transition regime between vibrational and
gamma-soft nuclei. By including additional copies of the algebra,
we can generate proton-neutron boson models involving other boson
degrees of freedom, while still maintaining exact solvability. In
each of these models, we can study not only the states of maximal
symmetry, but also those of mixed symmetry, albeit still in the
vibrational to gamma-soft transition regime. Furthermore, in each
of these models we can study some features of $F$-spin symmetry
breaking. We report systematic calculations as a function of the
pairing strength for models based on $s$, $d$, and $g$ bosons and
on $s$, $d$, and $f$ bosons. The formalism of exactly-solvable
models based on the SO(3,2) algebra is not limited to systems of
proton and neutron bosons, however, but can also be applied to
other scenarios that involve two species of interacting bosons.

\end{abstract}

 \maketitle

\section{Introduction}
\label{SecI}

In the early 1960s, Richardson \cite{Rich} showed how to exactly
solve  the pure pairing model including non-degenerate
single-particle orbits for both fermion \cite{Rich1} and boson
systems \cite{Rich2}. In 2001, Dukelsky, Esebbag and Schuck
\cite{DES} showed how to generalize Richardson's solution, making
use of analogous work by Gaudin \cite{Gaudin} for spin models, so
that the resulting Richardson-Gaudin models can now describe the
physics of a wide variety of strongly-correlated many-body quantum
systems governed by pairing correlations. Many interesting
applications of these exactly-solvable models have been reported
recently, to ultrasmall superconducting grains \cite{duke2}, to
nuclear superconductivity \cite{2DCoulomb}, to the nuclear
Interacting Boson Model \cite{FD,DP}, and to Bose Einstein
condensates \cite{DS}. The methods have been applied both to
fermion and boson systems, invariably yielding useful insight into
the properties of the complex quantum systems they model
\cite{Review}. Furthermore, with slight generalization the methods
have also proven useful in the description of mixed systems
involving atoms (either bosonic or fermionic) coupled to molecular
dimers in the presence of a Feshbach resonance \cite{DDEP}.

All of the above models were based on the treatment of pairing
correlations involving identical particles. The relevant algebraic
structure for describing identical-particle pairing is SU(2) for
fermions or SU(1,1) for bosons. Both are rank-one algebras with
just one Cartan generator. There has been several efforts to
derive exactly-solvable Richardson-Gaudin models based on larger
rank algebras \cite{Ush, Sie}. In particular, the rank-two SO(5)
algebra describing isovector proton-neutron pairing was treated in
\cite{Rich3}. More recently, the complete solution of the SO(5)
Richardson-Gaudin model was presented in ref.~\cite{so5}. This
made possible an exact treatment of isovector proton-neutron
pairing in the presence of non-degenerate single-particle levels,
which was used in a description of the $N=Z$ nucleus $^{64}$Ge in
an extended shell-model space.

In this paper, we describe the first example of an
exactly-solvable Richardson-Gaudin model of interacting {\it
bosons} that is based on a rank-two algebra. The model has the
algebraic structure SO(3,2) and is of relevance to two-component
boson systems. One possible realization is a mixture of $^{97}$Rb
atoms in the hyperfine states $|F=1, M_f=1\rangle$, $|F=1,
M_f=-1\rangle$ \cite{Myatt}. While these systems have been studied
\cite{Shi} in mean-field approximation, more elaborate approximate
methods or the use of exact solutions are needed to study effects
beyond mean field. Here we will concentrate on a different
realization, namely bosonic models of nuclei with distinct proton
and neutron degrees of freedom. When only $s$ and $d$ bosons are
included, the resulting models are restricted versions of the
proton-neutron Interacting Boson Model (IBM2) \cite{Ia}, in the
transitional regime between gamma-unstable and vibrational nuclei.
At the same time, however, their generality makes it possible to
address many issues that cannot be discussed within the context of
the IBM2. In particular, the models can readily accommodate boson
degrees of freedom other than just $s$ and $d$, still within a
proton-neutron framework. As such, they can address issues related
to the proton-neutron degree of freedom in boson models with $g$
bosons and/or $f$ bosons. The models  can also accommodate F-spin
breaking, which can thus likewise be studied in the presence of
bosons other than $s$ and $d$.

The outline of the paper is as follows. We begin in Section II
with a brief overview of the SO(3,2) algebra, confirming that it
is indeed a rank-two algebra and obtaining its associated
integrals of motion. In Section III, we show how this formalism
can be used to build exactly solvable models of relevance to
interacting neutron and proton bosons in atomic nuclei.  In
Section IV we consider as specific examples models involving $sd$,
$sdg$ and $sdf$ degrees of freedom and describe their exact
solutions in a variety of scenarios. In Section V, we summarize
the principal conclusions of this work.

\section{SO(3,2) and its Cartan decomposition}
\label{SecII}

In this section, we briefly review the elements of the Cartan
decomposition of the SO(3,2) Lie algebra and show how it can be
used to build the set of integrals of motion for models based on
multiple copies of this algebra.

We will be considering a problem involving neutron and proton
boson degrees of freedom, with creation and annihilation operators
$l^{\dagger}_{\rho m}$ and $l_{\rho m}$, respectively. These
operators create/annihilate a boson of angular momentum $l$,
projection $m$ and type $\rho$.

Now consider the set of bilinear boson operators
($\rho,~\rho'=\nu~or~\pi$)
\[
l^{\dagger}_{\rho} l_{\rho'}=\sum_m {l^{\dagger}_{\rho m} l_{\rho' m} }~,
\]

\[ b^{\dagger}_{-1l}= (-1)^{l/2} l^{\dagger}_{\nu} \cdot
l^{\dagger}_{\nu}~,\]

\[
b^{\dagger}_{0l}= (-1)^{l/2} \sqrt{2} l^{\dagger}_{\pi} \cdot
l^{\dagger}_{\nu}~,
\]

\begin{equation}
b^{\dagger}_{+1l}= (-1)^{l/2} l^{\dagger}_{\pi} \cdot
l^{\dagger}_{\pi}~,
\end{equation}
and the corresponding two-boson annihilation operators $b_{-1
1},~b_{01},~b_{+1 1}$. Here the scalar product $l^{\dagger}_{\rho}
\cdot l^{\dagger}_{\rho '}$ has the usual definition
\[
l^{\dagger}_{\rho} \cdot l^{\dagger}_{\rho '} = \sum_m (-)^{m}
l^{\dagger}_{\rho,m} l^{\dagger}_{\rho ',-m}~.
\]
For each $l$, there are four particle-hole operators , three pair
creation operators, and three pair annihilation operators. All
told, these ten operators generate an SO(3,2) algebra.

The Cartan decomposition of this algebra is achieved by rewriting
the set of one-body operators as

\[
H^1_l=F_{zl}={\frac 1 2} \left\{ \sum_m (l^{\dagger}_{\pi m} l_{\pi m} -
l^{\dagger}_{\nu m } l_{\nu m} ) \right\}~,
\]
\[
 H^2_l={\frac 1 2}\left(
 l^{\dagger}_{\pi} l_{\pi} + l^{\dagger}_{\nu} l_{\nu} +(2l+1)
 \right)~,
\]
\[
F^+_l = l^{\dagger}_{\pi} l_{\nu}= (-1)^l l^{\dagger}_{\pi}\cdot
\tilde{l}_{\nu}~,
\]
\begin{equation}
F^-_l = l^{\dagger}_{\nu} l_{\pi}=(-1)^l \ l^{\dagger}_{\nu}\cdot
\tilde{l}_{\pi}~.
\end{equation}

It can be readily confirmed that $[H_l^1,H_l^2]=0$. As there are
no other generators that simultaneously commute with both, SO(3,2)
is a rank-two algebra and these two operators generate the Cartan
subalgebra. [Note: the factor $1/2$ is included in the Cartan
operators to ensure that the Cartan-Weyl basis is orthonormal with
respect to the Killing form of ref.~\cite{ConformalField}.]

The remaining generators are the ladder operators of the algebra.
$F_l^{+}$ and $b^{\dagger}_{l \mu}$ are the raising operators and
correspondingly $F_l^-$ and $b_{l \mu}$ are the lowering
operators.

Knowing the Cartan decomposition for each $l$, we can obtain the
associated set of integrals of motion $R_l$, following
ref.~\cite{Sie}. The complete set of hermitean and mutually
commuting operators with linear and quadratic terms {\em in the
rational model} can be expressed in the form


\begin{equation}
R_{l}=\frac{\Delta}{2g}H^1_l+\frac{1}{2g}\left(2+\Delta\right)H^2_l+\sum_{l'(\neq
l)} \frac{X_l\cdot X_{l'}}{z_{l'}-z_{l}}~,
 \label{RI}
\end{equation}
where $H^{a}_l$ ($a=1,2$) are the two Cartan operators of the copy
$l$, and

\begin{equation}
X_l\cdot X_{l'}=F_{l}\cdot F_{l'}+H^2_{l}H^2_{l'} - \frac{1}{4}
\sum_{\mu =1,0,-1} (b^{+}_{\mu l}b_{\mu l'}+b^{+}_{\mu l'}b_{\mu l})~.
\end{equation}

For a problem involving $K$ copies of a rank-two SO(3,2) algebra
the integrals of motion depend on a total of $K+2$ free
parameters, the $K$ coefficients $z_l$ and the two parameters,
$\Delta$ and $g$, that enter in the linear term of eq.~(\ref{RI}).

The exact solutions of the rational model correspond to solving
the eigenvalue equations

\begin{equation}
R_l |\Psi(e_{\alpha},\omega_\gamma)\rangle = r_l
|\Psi(e_{\alpha},\omega_\gamma)\rangle~. \label{ansatz}
\end{equation}
The eigenvalues $r_l$ can be written as

\[
r_{l}=\frac{1}{2g}(U_{\pi l} + U_{\nu l}+\Delta U_{\pi
l})+\frac{1}{2g}(2+\Delta)\Omega_l
\]
\[
+  \frac{1}{2}\sum_{l'(\neq l)}
\frac{1}{z_{l'}-z_{l}}\Big(\Omega_{l}(U_{\pi l'}+U_{\nu l'}) +
\Omega_{l'}(U_{\pi l}+U_{\nu l})
\]
\[ +2\Omega_l\Omega_{l'}+U_{\pi l}U_{\pi l'}+U_{\nu
l}U_{\nu l'}\Big)
\]
\begin{equation}
+\frac{1}{2} (U_{\pi l}-U_{\nu l})\sum_{\gamma=1}^{M_1}
\frac{1}{\omega_{\gamma}-z_{l}} +(U_{\nu l}+\Omega_{l})
\sum_{\alpha=1}^{M_2}\frac{1}{e_{\alpha}-z_{l}}~,
\end{equation}
where $U_{\rho l} (\rho = \pi$ and $\nu)$ are the number of
non-paired $\rho$ bosons ({\em{i.e.}}, the seniority) in the level
$l$, and which define completely the  SO(3,2) irreducible
representation of copy $l$. Also, $\Omega_l=(2l+1)/2$ is
proportional to the degeneracy  of the levels,
$M_1=N_{\pi}-\sum_{l}U_{\pi l}$ is the total number of paired
$\pi$ bosons, and $M_2\equiv M=\frac{1}{2}(N_{\pi}-\sum_{l}U_{\pi
l}+N_{\nu}-\sum_{l}U_{\nu l})$ is the total number of pairs.
Finally, ${\omega_{\gamma}} ~ (\gamma=1,..,M_1)$, and
${e_{\alpha}}~ (\alpha=1,..,M_2)$ are two sets of spectral
parameters that must satisfy the set of coupled equations

\[
 \sum_{l} \frac{U_{\pi l}-U_{\nu l}}{z_{l}-\omega_{\gamma}} +
\sum_{\delta(\neq \gamma)}^{M_1} \frac{2}{
\omega_{\delta}-\omega_{\gamma} } - \sum_{\beta}^{M_2}
\frac{2}{e_{\beta}-\omega_{\gamma}} =-\frac{\Delta}{g} ~,
\]
\begin{equation}
\sum_{l} \frac{U_{\nu l}+\Omega_{l}}{z_{l}-e_{\alpha}} +
\sum_{\beta(\neq \alpha)}^{M_2} \frac{2}{ e_{\beta}-e_{\alpha} } -
\sum_{\delta}^{M_1} \frac{1}{\omega_{\delta}-e_{\alpha}} =-
\frac{1}{g} ~. \label{richeq}
\end {equation}
These are generalized Richardson equations appropriate to the
SO(3,2) algebra.

Once the free parameters that define the $R_l$ integrals of motion
have been chosen, any linear combination,
\begin{equation}
\sum_l \eta_l R_l ~,
\end{equation}
of those integrals of motion also gives rise to an
exactly-solvable model for the multiple copies of the SO(3,2)
algebra, with eigenvalues obtained from the corresponding $r_l$'s.
Of course, any constant can be added to the Hamiltonian without
affecting the exact solvability of the model.

For the scenario under discussion, a system of interacting neutron
and proton bosons in several levels, it is obviously desirable
that the exactly-solvable models that emerge have the appropriate
single-particle energies for the various boson degrees of freedom.
This is accomplished by requiring that the coefficients $\eta_l$
that define the linear combination of integrals of motion are
proportional to the single-particle energies of the corresponding
boson levels ($\eta_l=2g \epsilon_l$). If we furthermore choose
the $z_l$ parameters that enter in the definition of the integrals
of motion according to $z_l=2\epsilon_l$, the two-body part of the
model Hamiltonian likewise emerges in a very simple form, one in
which the interaction matrix elements are level-independent.
Furthermore, the constant term that is added to the Hamiltonian
can be chosen so as to leave it in a convenient form.

With these choices, the form of the Hamiltonian that emerges for
multiple copies of the SO(3,2) algebra is
\begin{eqnarray}
H&=&2g\sum_{l} \epsilon_l R_l+c \hat{1}
\nonumber\\
&=&\sum_l \epsilon_l\left( \hat{N}_{\pi l}+ \hat{N}_{\nu l}
 + \Delta  \hat{N}_{\pi l} \right) + \frac{g}{4} \sum_{l l' \mu}
b^{\dagger}_{\mu l} b_{\mu l'}
\nonumber\\
&&-\frac{g}{2} F \cdot F
+ \frac{g}{2}\left(B(U_\pi,U_\nu)+ C \right)\hat{1} ~,
\label{Hgeneral}
\end{eqnarray}
where
\begin{equation}
F=\sum_l F_l~,
\end{equation}
$c=g\sum_{l'>l}\Omega_l\Omega_{l'}-(2+\Delta)\sum_l\epsilon_l\Omega_l$
is a constant, and $\hat{1}$ is the identity operator.

 The diagonal term in the last line of eq.~(\ref{Hgeneral}) depends on
\begin{equation}
B(U_{\pi},U_{\nu})=\sum_{l} \Big( \frac{U_{\pi l}^2 +U_{\nu
l}^2}{2} + (U_{\pi l}+U_{\nu l})\Omega_{l} - U_{\nu l}-2U_{\pi l}
\Big)~,
\end{equation}
a function of the seniorities, and a constant $
C=N(4\Omega+6-N)/4$,  with $\Omega\equiv\sum_l \Omega_l$, and $N$
the total number of bosons ($N\equiv\sum_{l}N_{\pi l}+\sum_l
N_{\nu l} $).

The eigenvalues of the above Hamiltonian ($2g\sum_l\epsilon_l r_l+c$)
can be expressed as

\[
E =\sum_{l}\epsilon_{l}(U_{\pi l}+U_{\nu l} + \Delta U_{\pi l})
\]
\[
+ \sum_{\alpha}e_{\alpha}+\frac{g}{2}\left[\frac{\Delta}{g}
\sum_{\gamma}\omega_{\gamma}-(-F_z)(-F_z+1) \right] \]
\begin{equation}
 +\frac{g}{2}(B(U_\pi,U_\nu)+ C) ~,\label{Egeneral}
\end{equation}
where $F_z=(N_\pi-N_\nu)/2$.
 The  first line is the single
particle energy of the non-paired bosons. The first term of the
second line is the energy of the boson pairs, whereas the second
term is related to $F$-spin symmetry. It is possible to show that in
the $F$-spin-symmetric limit ($\Delta\rightarrow 0$), both terms in
the brackets combine to give:
$$\lim_{\Delta\rightarrow 0}
\left[\frac{\Delta}{g} \sum_{\gamma}\omega_{\gamma}-(-F_z)(-F_z+1)
\right] =-F(F+1)~.
$$

It is especially convenient to rewrite the Hamiltonian
(\ref{Hgeneral})
 in multipole-multipole form, to facilitate a link to the
traditional phenomenological Hamiltonians used in bosonic
descriptions of nuclei. The Hamiltonian (\ref{Hgeneral}) expressed
in terms of multipoles can be written as
\[
H=\sum_{l} \epsilon_l \left (\hat{N}_{\pi l} +\hat{N}_{\nu l} +
\Delta \hat{N}_{\pi l} \right)
\]
\begin{equation}
- \frac{g}{4} \sum_{l<l'}\sum_{L=|l'-l|}^{l'+l} (-)^L (
Q^L_{l_{\nu}l'_{\nu}}+Q^L_{l_{\pi}l'_{\pi}}  ) \cdot (
Q^L_{l'_{\nu}l_{\nu}}+Q^L_{l'_{\pi}l_{\pi}}  ) ~,\label{Hll}
\end{equation}
where
\begin{equation}
Q^L_{l_{\rho}l'_{\rho}} = \left( l^{\dagger}_{\rho}
\tilde{l'}_{\rho} - (-)^{l+\frac{l+l'}{2}} l'^{\dagger}_{\rho}
\tilde{l}_{\rho} \right)^{L}~.\label{Qll}
\end{equation}





\section{SO(3,2) and the proton-neutron Interacting Boson
Model} \label{SecIII}

The multi-copy SO(3,2) formalism described in the preceding
section has relevance to the proton-neutron Interacting Boson
Model of atomic nuclei \cite{Ia}. In this model, distinct neutron
and proton bosons are introduced to model collective pairs of
identical nucleons. The standard version of the model is limited
to $s$ and $d$ bosons, reflecting the fact that the energetically
lowest pairs of identical nucleons are invariably those with
$J^P=0^+$ and $J^P=2^+$ and that they are typically
well separated from all higher pairs. The resulting model, usually
called the IBM2, has been used to describe successfully the
collective properties of nuclei throughout the periodic table.

Soon after the introduction of the IBM2, it became clear that
there was a natural connection of the model both to the nuclear
shell model and to the simpler IBM1 model with but one kind of
boson.

The relationship to the shell model is contained in the
association of the bosons with the lowest collective pairs of
identical nucleons. Thus, it is possible to use input from the
shell model to help define the Hamiltonian that acts in the IBM2
model space.

The relationship to the IBM1 comes through the introduction of a
quantum number called $F$ spin that distinguishes the two species in
IBM2, the neutron and proton bosons. In an IBM2 Hamiltonian that
conserves this quantum number, the lowest states of the system are
those that are maximally symmetric in $F$-spin, in other words
maximally symmetric under the interchange of neutron and proton
bosons. It is these states that are modelled by the simpler IBM1.
However, the IBM2 also contains states that are not maximally symmetric
under the interchange of neutron and proton bosons, states
that are said to have mixed symmetry. In an $F$-spin conserving
version of IBM2, these states are decoupled from the states of
maximal $F$-spin symmetry.

With the above as background, we are now in a position to discuss
the relationship between the exactly solvable models that derive
from the SO(3,2)algebra and the IBM2. In particular, if we
limit our discussion to two copies of the algebra, one involving
an $l=0$ $s$ boson and the other involving an $l=2$ $d$ boson,
then we arrive precisely at a Hamiltonian of the IBM2 form. It is,
however, not the most general IBM2 Hamiltonian. Rather, as we will
see shortly it is a Hamiltonian that models the transition between
vibrational (U(5)) and gamma-soft (SO(6)) nuclei within an IBM2
framework.

The Hamiltonian that derives in a two-copy $s$-$d$ realization of
SO(3,2) has the form (see eqs.~(\ref{Hll},\ref{Qll}), with
$l,l'=s,d$ only, and taking $\epsilon_s=0$)
\[
H=\epsilon_d \left(\hat{N}_{\pi d} +\hat{N}_{\nu d} + \Delta
\hat{N}_{\pi d} \right)
\]
\begin{equation}
- \frac{g}{4} (\hat{Q}^2_{\pi} +\hat{Q}^2_{\nu}) \cdot
(\hat{Q}^2_{\pi} +\hat{Q}^2_{\nu})~, \label{Hsd}
\end{equation}
where
\begin{equation}
\hat{Q}^2_{\rho} = s^{\dagger}_{\rho} \tilde{d}
+d^{\dagger}_{\rho} \tilde{s}_{\rho} ~. \label{Qsd}
\end{equation}
Its eigenvalues can be obtained from eq. (\ref{Egeneral}).

There are a couple of features of this Hamiltonian worthy of note.
The first is that as mentioned above it is not a completely
general IBM2 Hamiltonian, as the quadrupole operator is restricted
to $\chi=0$. Second, the Hamiltonian contains a term $\Delta
\hat{N}_{\pi d}$, which arises if the single-boson splittings
between the $s$ and $d$ levels in the two species (neutrons and
protons) are different, and which will in general break $F$-spin
symmetry. $F$-spin symmetry is only preserved if $\Delta=0$.
Finally, the quadrupole-quadrupole interaction that enters is an
$F=0$ tensor, with a definite relation between the interactions
among the different species. There is no additional Majorana
interaction in the model, as the term $g F \cdot F$ in eq.
(\ref{Hgeneral}) is absorbed completely into the $F$-spin-scalar
quadrupole-quadrupole interaction. However, when $\Delta=0$, {\em
i.e.} when $F$-spin symmetry is preserved, we can add an additional
Majorana term of arbitrary strength while maintaining the exact
solvability of the model.

Perhaps the key feature of the exactly-solvable proton-neutron
boson models that derive from the multi-copy SO(3,2) algebra is
that they can accommodate boson degrees of freedom other than just
the $s$ and $d$. Indeed, they can accommodate any number of boson
degrees of freedom, albeit with the restricted Hamiltonians that
can  be accessed. As an example, the model with $s,~ d$ and $g$
degrees of freedom is of the form ($\epsilon_s=0$)
\[
H=\sum_{l=d,g} \epsilon_l \left(\hat{N}_{\pi l} +\hat{N}_{\nu l} +
\Delta \hat{N}_{\pi l} \right)
\]
\begin{equation}
- \frac{g}{4} \sum_{l<l'}\sum_{L=2,...,6} (-)^L (
Q^L_{l_{\nu}l'_{\nu}}+Q^L_{l_{\pi}l'_{\pi}}  ) \cdot (
Q^L_{l'_{\nu}l_{\nu}}+Q^L_{l'_{\pi}l_{\pi}}  ) ~.\label{Hsdg}
\end{equation}
Note that it contains all multipole interactions from $L=2$ to
$L=6$. Once again, all enter as $F=0$ tensors, with no additional
Majorana interaction present. Here too, however, a Majorana
interaction of arbitrary strength can be included if $\Delta=0$ so
that $F$-spin is conserved.

Likewise, we can accommodate $f$ bosons, either with or without
$g$ bosons, still in a proton-neutron boson framework. When
considering $sd$ and $f$ bosons, for example, the class of
Hamiltonians that arise are ($\epsilon_s=0$)

\[
H=\sum_{l=d,f} \epsilon_l \left(\hat{N}_{\pi l} +\hat{N}_{\nu l} +
\Delta \hat{N}_{\pi l} \right)
\]
\begin{equation}
- \frac{g}{4} \sum_{l<l'}\sum_{L=1,...,5} (-)^L (
Q^L_{l_{\nu}l'_{\nu}}+Q^L_{l_{\pi}l'_{\pi}}  ) \cdot (
Q^L_{l'_{\nu}l_{\nu}}+Q^L_{l'_{\pi}l_{\pi}}  ) ~.\label{Hsdf}
\end{equation}
Note that now we only have multipoles from $L=1$ through $5$. The
$L=1,~3,~4$ and $5$ multipoles all have odd parity;  the $L=2$
multipole has an even-parity component ($l,l'=s,d$) and an
odd-parity component ($l,l'=d,f$).

\section{Applications of exactly-solvable proton-neutron interacting boson
models derived from the SO(3,2) algebra}
\label{SecIV}

In this section, we apply the various exactly-solvable model
Hamiltonians developed in the previous section to scenarios of
relevance to the collective structure of atomic nuclei.

We first consider a model of $s$, $d$ and $g$ interacting neutron
and proton bosons, for which the relevant Hamiltonian was given in
eq.~(\ref{Hsdg}). We compare the results for this model with those
of the corresponding model with just $s$ and $d$ bosons, (see
eq.~(\ref{Hsd})). The analysis is carried out as a function of the
pairing strength $g$, for fixed single-boson energies, with the
results shown in fig. 1.

The complete Hilbert space of the model can be split into
invariant subspaces characterized by the seniority numbers
($U_{\pi l}, U_{\nu l}$). For each seniority, one has to solve a
different set of Richardson equations (\ref{richeq}) to obtain the
eigenvalues (\ref{Egeneral}). The angular momenta, parities, and
F-spin quantum numbers associated with the set of seniorities we
consider are shown in table \ref{seniorities}.

\begin{table}
\begin{tabular}{|c|c|c|c|}
\hline
& & \\
$(\scriptstyle {U_{\pi s} U_{\nu s} U_{\pi d} U_{\nu d} U_{\pi g}
U_{\nu g} } )$&
$F$  & $J^P$\\
  & & 
  \\
 \hline
(0 0 0 0 0 0)& $0,1,\ldots,F_{max}$& $0^+$ \\
\hline
(0 1 0 1 0 0)& $0,1,\ldots,F_{max}$& $2^+$\\
(0 1 0 0 0 1)& $0,1,\ldots,F_{max}$& $4^+$\\
(0 0 0 2 0 0)& $0,1,\ldots,F_{max}$& $2^+,4^+$\\
(0 0 1 1 0 0)& $0,1,\ldots,F_{max}-1$& $1^+,3^+$\\
(0 0 0 1 0 1)& $0,1,\ldots,F_{max}$& $2^+,3^+,4^+,5^+,6^+$\\
(0 0 0 0 0 2)&$0,1,\ldots,F_{max}$ & $2^+,4^+,6^+,8^+$\\
(0 0 0 0 1 1)& $0,1,\ldots,F_{max}-1$& $1^+,3^+,5^+,7^+$\\
\hline
(0 1 0 3 0 0)& $0,1,\ldots,F_{max}$& $0^+,3^+,4^+,6^+$\\
(0 0 0 4 0 0)&$0,1,\ldots,F_{max}$ &$2^+,4^+,5^+,6^+,8^+$ \\
(0 1 0 2 0 1)& $0,1,\ldots,F_{max}$ & $0^+,1^+,\ldots,8^+$ \\
(0 1 0 1 0 2)&$0,1,\ldots,F_{max}$ & $0^+,1^+,\ldots,10^+$\\
\hline
\end{tabular}
\caption{Seniorities considered in the sd and sdg calculations and
their associated angular momenta, parities and F-spin quantum
numbers.} \label{seniorities}
\end{table}

\begin{figure}
 \includegraphics[height=.40\textheight]{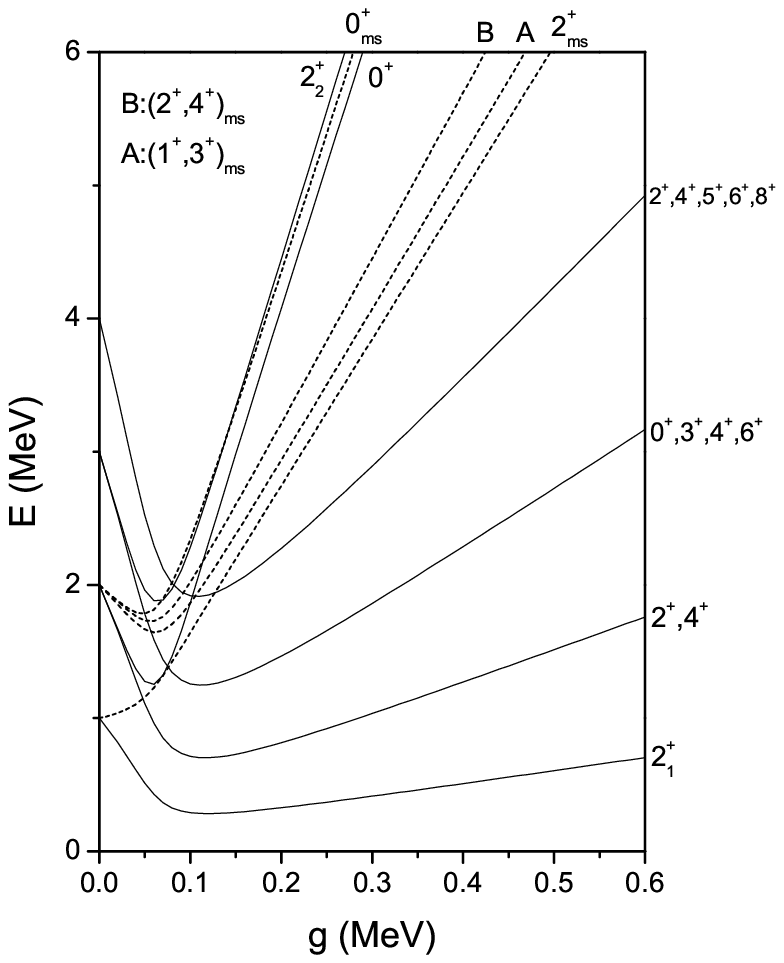}
\includegraphics[height=.40\textheight]{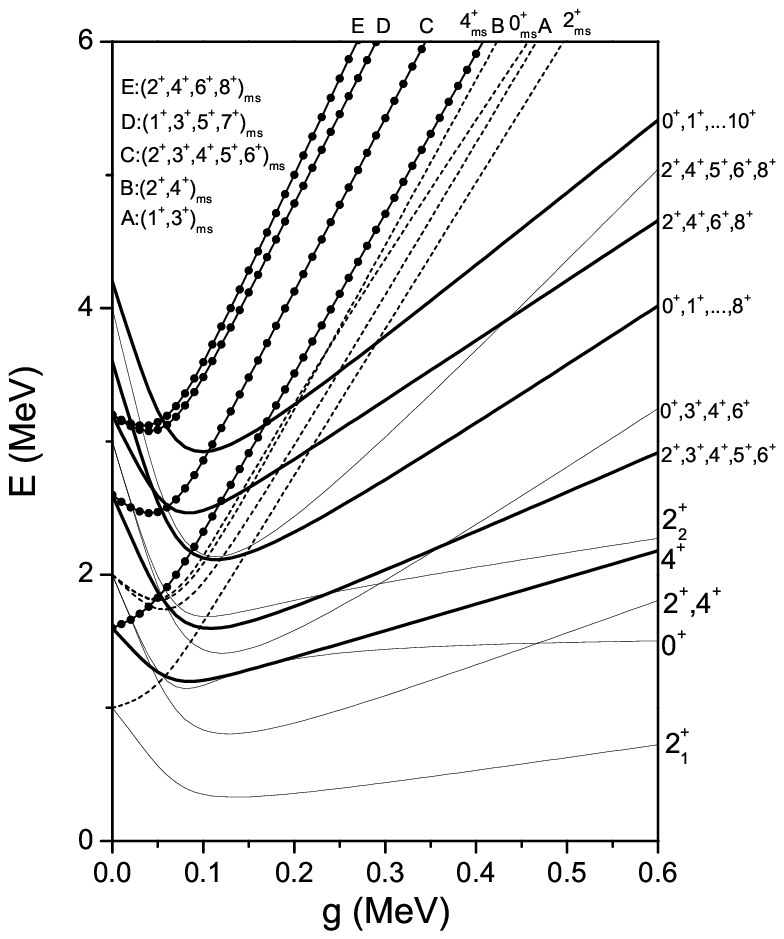}
  \caption{Excitation energies  of the $sdg$ Hamiltonian in eq. (\ref{Hsdg}) (bottom panel) compared to those of the
 $sd$ Hamiltonian in eq. (\ref{Hsd}) (top panel), as a function of the interaction strength $g$ and for
 the seniorities of table \ref{seniorities}.
 Continuous lines correspond
 to maximal F-spin states,  dashed and circled  lines refer to
 mixed symmetry states ($F=F_{\hbox{max}}-1$). Thick  and circled
 lines in the bottom panel indicate states with
 non-paired g-bosons ($U_g\not=0$),  for maximal and mixed symmetry states respectively.
 Except for $J^P=2^+$, only the
 first excited state  of each  seniority is shown.  Only those mixed
 symmetry states
 up to total seniority two are presented.  The single-boson energies are given in the text.
}
\end{figure}
The top panel gives the results for the $sd$ model and the bottom
panel gives the $sdg$ results. All energies are with respect to
that of the ground state.

In the case of the $sd$ model, we use $\epsilon_d=1~MeV$ and
$\epsilon_s=0~MeV$, and the relevant seniorities are those with
$U_g=0$ in table \ref{seniorities}. For the $sdg$ model, we again
use $\epsilon_d=1~MeV$, and $\epsilon_s=0~MeV$, but now include a
$g$ boson at $\epsilon_g=1.6~MeV$. In this case, we consider all
the seniorities of table \ref{seniorities}. The level energies
chosen are physically reasonable for the various boson degrees of
freedom. In both calculations, we consider a system of $N=20$
bosons, with $N_{\nu}=N_{\pi}=10$ ($F_{max}=10$). Furthermore, we
assume $\Delta=0$ so that $F$-spin symmetry is preserved.

Note that $g=0$ corresponds to the precise $U(5)$ limit in the
$sd$ model and to the corresponding vibrational limit in the $sdg$
model. As the pairing strength $g$ increases, there is a phase
transition to a gamma-unstable system, occurring at roughly
$g=0.08$~MeV.

Some interesting features can be noted in these results. Inclusion
of the $g$ level in the calculation increases significantly the
number of possible $J^P$ states (indicated by thick and circled
lines in the bottom panel of fig. 1, see table \ref{seniorities}),
but otherwise leaves most of the states of the $sd$ model
relatively unaffected. For a few states, however,  and in
particular the levels denoted $0^+$, $0^+_{\hbox{ms} }$, and
$2_2^+$, there are important and interesting changes that take
place. As an example, consider the first excited $0^+$ state.  In
the $sd$ calculation, its energy goes up as $g$ increases; in the
corresponding $sdg$ calculation its energy flattens out.  A
similar effect can be seen for the $0^+_{\hbox{ms} }$ and $2_2^+$
states, which likewise are flattened out in energy.

To better illustrate this phenomenon and how it arises, we present
in the bottom panel of fig.~2 the pair energies ($e_\alpha$)
associated with some selected seniority-zero states ($J^P=0^+$) of
the $sdg$ model, as a function of the interaction strength. We
also show the associated spectrum of levels in the top panel of
the figure.

For the ground state of the system (GS), all pair energies are
trapped between $2\epsilon_d$ and $2\epsilon_s$. Then there are a
series of levels in which one or more of the pair energies are
promoted into the energy region from $2\epsilon_d$ to
$2\epsilon_g$. The corresponding levels are likewise trapped, as
their energies are simply a sum of the associated pair energies.
The lowest is $\alpha_1$, with but one pair excited. The next is
$\alpha_2$, with two pairs excited. And as should be evident there
are many other such states, in which at least one pair is between
$2\epsilon_g$ and $2\epsilon_d$, but in which none are above
$2\epsilon_g$. For simplicity, however, we only show the lowest
two levels of this {\em band}.

Next we consider levels in which one pair is excited above
$2\epsilon_g$. The lowest such level is denoted $\beta_1$ in both
panels. Since the last pair is not trapped, this level goes up in
energy as $g$ increases. Furthermore, there will be a band of
levels in which just one pair has been excited above
$2\epsilon_g$. Once again, however, we only show the two lowest
states of the band.

Finally, we show two other bands in the spectrum, denoted $\gamma$
and $\delta$. They correspond to two and three pairs above
$2\epsilon_g$, respectively.  They too are unbounded as a function
of $g$. Of course there are higher bands as well, corresponding to
progressively more pairs being excited above $2\epsilon_g$.

As should be clear from the figures and the discussion, this
unusual behavior for selected levels occurs whenever there are
more than two boson levels active and a repulsive boson-boson
pairing interaction \cite{DP}.

\begin{figure}
  \includegraphics[height=.40\textheight]{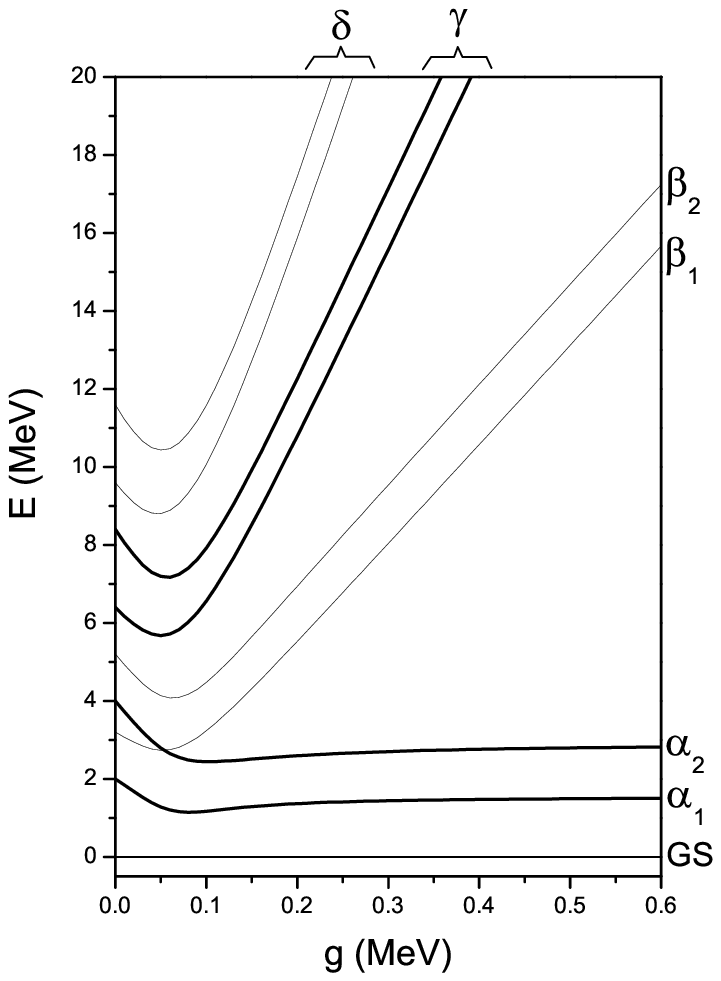}
  \includegraphics[height=.40\textheight]{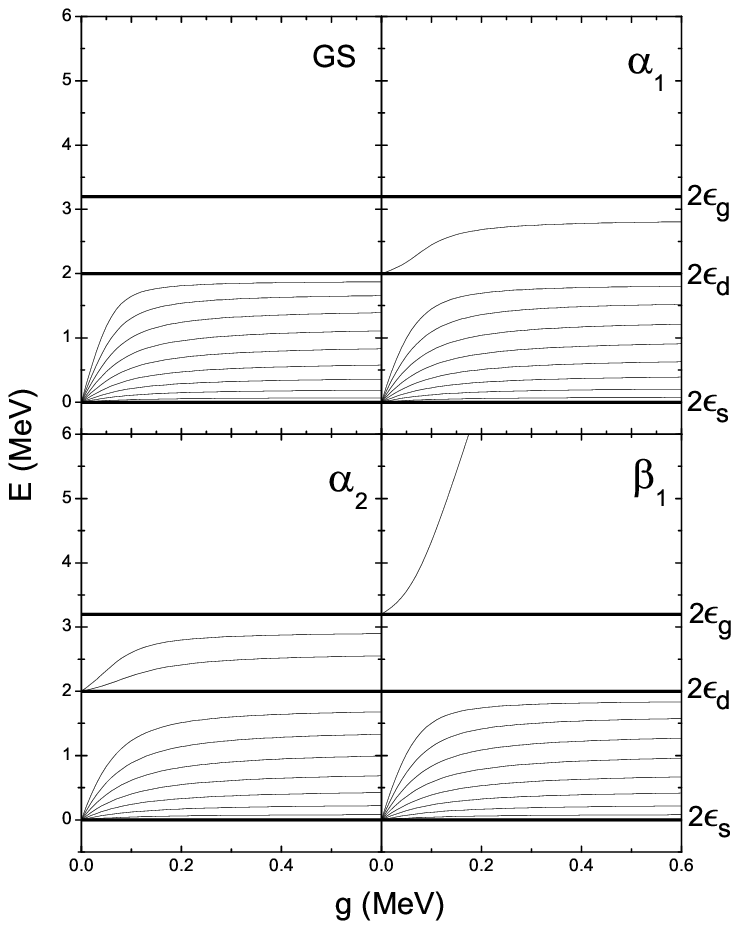}
   \caption{Excitation energies of the $sdg$ Hamiltonian in eq. (\ref{Hsdg}) (top panel) for
  seniority-zero states ($J^P=0^+$). The thickness of the lines serves as a guide to the eye to
  distinguish the band structure of the spectrum ($\alpha,~\beta,~\gamma~,\delta$).
    Only the first two states for each band are plotted. In the bottom panel the pair energies
    ($e_\alpha, \alpha=1,...,10$) associated with
    the ground state (GS), the first ($\alpha_1$) and second ($\alpha_2$) states of the lowest excited
    band, and the first ($\beta_1$) state of the second excited band
    are shown (see text for details). All
    the parameters are the same as for the bottom panel of fig.~1.
  }
\end{figure}

As noted earlier, the trapping of pair energies is not restricted
to $0^+$ states. Indeed, the band structure found for
seniority-zero states is a feature common to all seniority
sectors. However, for the states with seniority different from
zero the term $(g/2)B(U_\pi,U_\nu)$ in the eigenvalues
(\ref{Egeneral}) also contributes, making the excitation energies
for such states unbounded as a function of the interaction
strength $g$. Nevertheless, these states likewise arise from
trapped pair energies and thus are also flattened out as a
function of $g$.

For the mixed-symmetry states in the $\Delta=0$ limit, the term in
brackets in (\ref{Egeneral}) produces a  contribution to the excitation energy,
$$
\frac{g}{2}\left(F_{\rm max}(F_{\rm max}+1)-F(F+1)\right),
$$
thereby explaining why they appear at higher energy than the
maximal $F$-spin states. As was mentioned above, however, when
$\Delta=0$ it is possible to add a Majorana term of arbitrary
strength, and thus move the location of the mixed-symmetry states
at all with respect to those of maximal $F$-spin.

We next consider the effect of including an $f$ boson, focussing
now on an $sdf$ boson model. In fig.~3, we present the results for
this model as a function of $g$, for the case of
$\epsilon_s=0$~MeV, $\epsilon_d=1$~MeV and $\epsilon_f=1.6$~MeV.
The seniorities that were considered in this calculation are
listed in table \ref{senioritief}.

\begin{table}
\begin{tabular}{|c|c|c|c|}
\hline
& & \\
$(\scriptstyle {U_{\pi s} U_{\nu s} U_{\pi d} U_{\nu d} U_{\pi f}
U_{\nu f} })$&
$F$  & $J^P$\\
  & & 
  \\
 \hline
(0 0 0 0 0 0)& $0,1,\ldots,F_{max}$& $0^+$ \\
\hline
(0 1 0 1 0 0)& $0,1,\ldots,F_{max}$& $2^+$\\
(0 1 0 0 0 1)& $0,1,\ldots,F_{max}$& $3^-$\\
(0 0 0 2 0 0)& $0,1,\ldots,F_{max}$& $2^+,4^+$\\
(0 0 1 1 0 0)& $0,1,\ldots,F_{max}-1$& $1^+,3^+$\\
(0 0 0 1 0 1)& $0,1,\ldots,F_{max}$& $1^-,2^-,3^-,4^-,5^-$\\
(0 0 0 0 0 2)&$0,1,\ldots,F_{max}$ & $2^+,4^+,6^+$\\
(0 0 0 0 1 1)& $0,1,\ldots,F_{max}-1$& $1^+,3^+,5^+$\\
\hline
(0 1 0 3 0 0)& $0,1,\ldots,F_{max}$& $0^+,3^+,4^+,6^+$\\
(0 0 0 4 0 0)&$0,1,\ldots,F_{max}$ &$2^+,4^+,5^+,6^+,8^+$ \\
(0 1 0 2 0 1)& $0,1,\ldots,F_{max}$ & $1^-,2^-,\ldots,7^-$ \\
(0 1 0 1 0 2)&$0,1,\ldots,F_{max}$ & $0^+,1^+,\ldots,8^+$\\
\hline
\end{tabular}
\caption{Seniorities considered in the $sdf$ calculations and
their associated angular momenta, parities  and F-spin quantum
numbers.} \label{senioritief}
\end{table}

Thus, this model is very similar to the $sdg$ model shown in
fig.~1, with the $f$ boson replacing the $g$ boson at precisely
the same energy.  Here too the second $0^+$ level is trapped, as
this is a general phenomenon that occurs whenever there are more
than two boson degrees of freedom and a repulsive pairing
interaction. Indeed, the spectra for the $sdf$ and $sdg$ models as
a function of $g$ are very similar. The exactly-solvable SO(3,2)
Richardson-Gaudin models do not couple states with broken pairs,
which is where differences would have shown up.

\begin{figure}
  \includegraphics[height=.40\textheight]{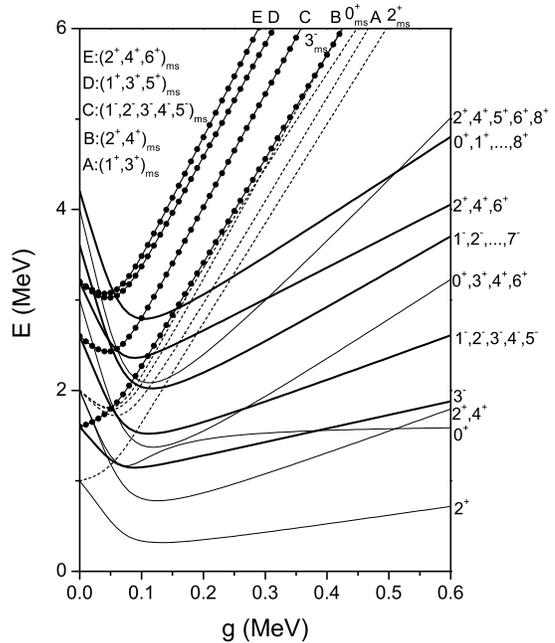}
 \caption{Excitation energies of the $sdf$ Hamiltonian in eq.~(\ref{Hsdf})  as a function of the interaction strength
  $g$, for the seniorities of table \ref{senioritief}. Continuous lines correspond
 to maximal F-spin states,  dashed and circled  lines refer to mixed symmetry states ($F=F_{max}-1$).
 Thick  and circled lines  indicate states with
  non-paired f-bosons ($U_f\not=0$),  for maximal and mixed F-symmetry states
  respectively. Only the first excited sate of each seniority is
  shown. For mixed symmetry states,  only  states  up to total seniority two are
  plotted.
  The single-boson energies are given in the text.}
\end{figure}

\section{Summary and Conclusions} \label{SecV}

In this paper, we have described the first example of an
exactly-solvable Richardson-Gaudin model based on a rank-two
algebra for bosons. The example we discussed involved multiple
copies of the SO(3,2) algebra. We focussed on a specific
realization involving proton and neutron bosons subject to a
pairing interaction. The models that emerged are exactly solvable
for non-degenerate single-boson energies and for any number of
bosonic copies. When only $s$ and $d$ bosons are included, the
resulting model is a specific example of the proton-neutron
Interacting Boson Model (IBM2), limited however to the transition
from vibrational to gamma-soft motion. Other bosonic degrees of
freedom, such as the $g$ and/or $f$, can be readily included while
still preserving the exact solvability of the model. Through the
use of these models, we can address issues related to the role of
mixed-symmetry states in proton-neutron boson models with many
boson degrees of freedom. We can even study some limited features
associated with $F$-spin symmetry breaking for these multi-copy
boson models.

One issue addressed here concerned the role of the $g$ boson in
proton-neutron boson models. We compared the results of $sd$
calculations with those of $sdg$ calculations for an
$F$-spin-symmetric Hamiltonian, throughout the vibrational to
gamma-soft transition region, and saw that in the presence of a
$g$ boson several low-lying excited states have their properties
dramatically modified as the result of a trapping phenomenon for
the pair energies.

We also studied the role of the $f$ boson, in the context of an
$sdf$ model of interacting proton and neutron bosons. Because of
some special features of the exactly-solvable models that we
treat, there were no meaningful differences between the $sdg$ and
$sdf$ results, except of course as regards the angular momentum
content of the collective states.

The models that we have developed are not limited, however, to
proton-neutron boson models of nuclei. Any physical problem
involving two species of bosons in which pairing is dominant can
be modelled in this way. In Section I, we noted the possibility of
applying these models to problems involving  a mixture of
$^{97}$Rb atoms in the hyperfine states $|F=1, M_f=1\rangle$,
$|F=1,M_f=-1\rangle$.
The availability of an exactly-solvable model with which to
study such systems, makes it possible to study effects that go
beyond the limited mean-field descriptions considered to date.

\begin{flushleft}
{\bf Acknowledgements}
\end{flushleft}
This work was supported in part by the Spanish DGI under grant No.
BFM2003-05316-C02-02, by the US National Science Foundation under
grant \# PHY-0140036, and by a CICYT-IN2P3 cooperation. SLH has a
post-doctoral fellowship from the Spanish SEUI-MEC and BE has a
pre-doctoral grant from the Spanish CE-CAM. SP would like to
acknowledge the hospitality and support of the CSIC, Madrid, where
most of this work was carried out.

\end{document}